\documentclass[aps,prl,showpacs,twocolumn]{revtex4}

\usepackage{amsmath,amssymb,graphicx}

\newcommand{\be}{\begin{equation}}
\newcommand{\ee}{\end{equation}}
\newcommand{\bea}{\begin{eqnarray}}
\newcommand{\eea}{\end{eqnarray}}

\begin{document}

\noindent{\bf Comments on ``Origin of cosmic magnetic fields''}\\


\noindent In a recent Letter \cite{Campanelli:2013mea} it has been proposed that primordial magnetic fields are generated during a de Sitter inflationary phase, contrarily to what is commonly believed, and that they are the seeds of the galactic and galaxy cluster magnetic fields observed today. The key result  is that the two-point function of the magnetic field, once renormalized with adiabatic subtraction, is constant and proportional to $H^{4}$ in de Sitter space-time. In this Comment, we show that, although the result is mathematically correct (it is simply a calculation of the conformal anomaly in de Sitter spacetime), it does not have any physical relevance  because inflation cannot be modeled by a de Sitter phase without beginning.

In a realistic model of inflation, at the beginning of the quasi-exponential expansion all physically  relevant modes are sub-Hubble modes, short wavelength modes "inside the horizon". The very large scale modes  which are already outside the horizon initially are not physically relevant.  However, these last ones are precisely the modes  responsible for the result of \cite{Campanelli:2013mea} and this is the main reason why we believe that the result is unphysical. As we show below, if one introduces some maximal wavelength, $\lambda_{\max}=2\pi/k_{\min}$ with $k_{\min}\neq 0$, the entire renormalized two-point function vanishes.

Let us  substantiate our claim. We follow Ref.~\cite{Campanelli:2013mea} (see the Letter for details), and introduce the  ``physical'' spectrum ${\cal P}_{phys}(k, m)={\cal P}(k, m)-{\cal P}^{(A)}(k,m)$, where ${\cal P}^{(A)}(k,m)$ is the adiabatic expansion up to fourth order \cite{PTbook}, 
and $m$ the photon regulator mass. This removes the ultraviolet divergence of the bare 2-point function and  leads to the result shown in Figure 1. One immediately notes that this cannot be the power spectrum of a well-defined correlation function since it becomes negative at $k/a \sim m$. This signifies, in the best case, that the correlation function has singularities. This is not surprising since the adiabatic subtraction scheme employed here is valid only for $k/a\gg m$. A similar behavior has also been observed in Ref.~\cite{Perrier:2012nr} in the regime where the adiabatic subtraction is not valid. 

Figure 1 shows that the main contribution to the 2-point function comes  from long wave modes. In particular, in the vanishing mass limit, ${\cal P}_{phys}$  becomes proportional to $\delta(k)$ as pointed out in \cite{Campanelli:2013mea}.  If we now introduce a maximal wavelength, $k_{\min}>0$,
and compute the magnetic field from modes above this cutoff, we obtain  \vspace{-0.3cm}
$$\langle {\bf B}(\vec{x}, t)^2 \rangle=\lim_{m\rightarrow 0}\int_{k_{\min}}^{\infty} dk k^{-1} {\cal P}_{phys}(k,m) =0, \vspace{-0.3cm}$$ independent of the value of $k_{\min}$.

This means that, by discarding the modes that are outside the horizon at any fixed beginning of the de Sitter inflationary phase, the amplification of the magnetic field vanishes.  
Hence, the adiabatic subtraction affects significantly the infrared tail of the physical power spectrum.

On the other hand, in  a realistic  inflationary scenario, where quantum fluctuations take on a time-dependent effective mass, adiabatic subtraction does not alter the spectrum of infrared modes with $k/a\ll m (\ll H)$ \cite{Durrer:2009ii,Marozzi:2011da,Finelli:2007fr}. Let us just remark here that in Refs.~\cite{Durrer:2009ii,Marozzi:2011da} we do not introduce any infrared cutoff, hence the criticism in footnote [13] of \cite{Campanelli:2013mea} does not apply. 

In conclusion, we believe that the result in \cite{Campanelli:2013mea} is unphysical, 
and that pushing the adiabatic subtraction into the far-infrared regime can lead to pathological results in  de Sitter space.

\begin{figure}[th]
\centering
\includegraphics[width=8cm]{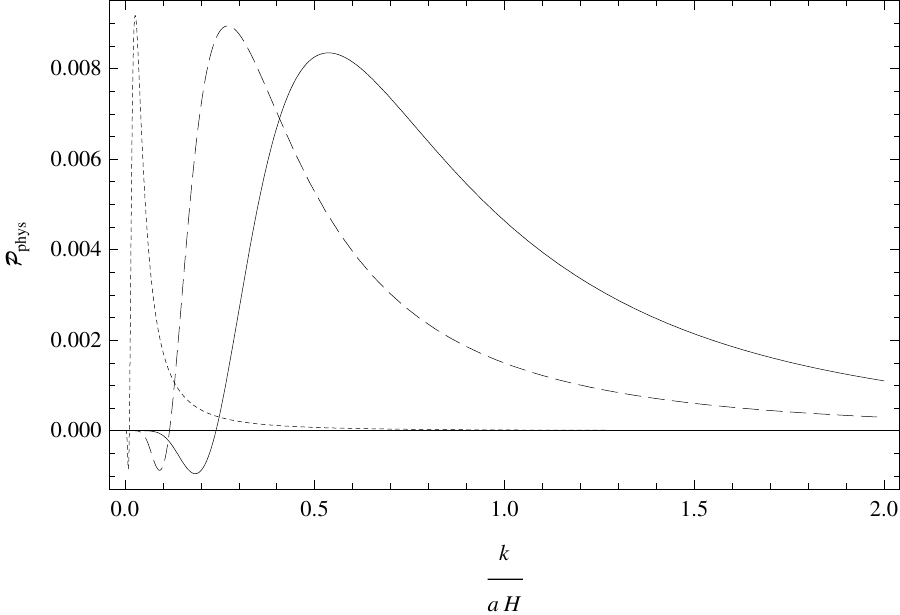}
\centering
\caption{The physical spectrum ${\cal P}_{phys}$ is plotted in units of $H^4$ as a function of $k/(a H)$ for different values of $m/H$
($m/H=1/5$ solid line, $m/H=1/10$ dashed line, and $m/H=1/100$ dotted line).}
 \label{f1}
\end{figure}

\noindent {\it Acknowledgement.} 
We thank J.\ Cline  and M.\ Peloso for useful discussions and comments.
G.M. is supported by the Marie Curie IEF, Project NeBRiC.  
M.R. is supported by the ARC convention No.11/15-040. R.D. acknowledges the Swiss National Science Foundation.\\


\vspace{1mm}

\noindent Ruth Durrer$^1$, Giovanni Marozzi$^1$, and Massimiliano Rinaldi$^2$\\
$^1$Universit\'e de Gen\`eve, DPT and CAP, 
24 quai Ernest-Ansermet, CH-1211 Gen\`eve 4, Switzerland\\
$^2$Namur Center for Complex Systems and University of Namur, 8 Rempart de la Vierge, B-5000, Belgium


\end{document}